\documentclass{article}
\usepackage{spconf,amsmath,epsfig}

\title{EXCITATION-BASED VOICE QUALITY ANALYSIS AND MODIFICATION}
\twoauthors
  {Thomas Drugman, Thierry Dutoit}
	{Facult\'e Polytechnique de Mons\\
	TCTS Lab\\
	Belgium}
  {Baris Bozkurt}
	{Izmir Institute of Technology\\
	Dpt. of Electrical \& Electronics Engineering\\
	Turkey}

\begin{document}
%\ninept
%
\maketitle
\begin{abstract}
This paper investigates the differences occuring in the excitation for different voice qualities. Its goal is two-fold. First a large corpus containing three voice qualities (modal, soft and loud) uttered by the same speaker is analyzed and significant differences in characteristics extracted from the excitation are observed. Secondly rules of modification derived from the analysis are used to build a voice quality transformation system applied as a post-process to HMM-based speech synthesis. The system is shown to effectively achieve the transformations while maintaining the delivered quality.
\end{abstract}

%%%%%%%%%%
%%%%%%%%%%
\section{Introduction}\label{sec:intro}

Since early times of computer-based speech synthesis research, voice quality (the perceived timbre of speech) analysis/modification has attracted interest of researchers \cite{Klatt}. The topic of voice quality analysis finds application in various areas of speech processing such as high-quality parametric speech synthesis, expressive/emotional speech synthesis, speaker identification, emotion recognition, prosody analysis, speech therapy. Due to availability of reviews such as \cite{Alessandro} and space limitations, a review of voice quality analysis methods will not be presented here.

For voice quality analysis of speech corpora, it is common practice to estimate spectral parameters directly from speech signals such as relative harmonic amplitudes, or Harmonic to Noise Ratio (HNR). Although the voice quality variations are mainly considered to be controlled by the glottal source, glottal source estimation is considered to be problematic and hence avoided in the parameter estimation procedures for processing large speech corpora. In this work, we follow the not so common path and study the differences present in the glottal source signal parameters estimated via an automatic algorithm when a given speaker produces different voice qualities. Based on a parametric analysis of these latter (Section \ref{sec:VQanalysis}), we further investigate the use of the information extracted from a large corpus, for voice quality modification of other speech databases in a HMM-based speech synthesizer (Section \ref{sec:VQmodif}).

%%%%%%%%%%
%%%%%%%%%%
\section{Excitation-based Voice quality analysis}\label{sec:VQanalysis}

The goal of this part is to highlight the differences present in the excitation when a given speaker produces different voice qualities. The De7 database used for this study was designed by Marc Schroeder as one of the first attempts of creating diphone databases for expressive speech synthesis \cite{Schroder}. The database contains three voice qualities (modal, soft and loud) uttered by a German female speaker, with about 50 minutes of speech available for each voice quality. In Section \ref{ssec:glottal}, the glottal flow estimation method and glottal flow parametrization used in this work are briefly presented. The harmonicity of speech is studied via the maximum voiced frequency in Section \ref{ssec:Fm}. As an important perceptual charactersitic, spectral tilt is analyzed in Section \ref{ssec:SpectralTilt}. Section \ref{ssec:Eigen} compares the so-called \emph{eigenresiduals} \cite{DSM} of the different voice qualities. Finally Section \ref{ssec:Separability} quantifies the separability between the three voice qualities for the extracted excitation features.

% For this, the De7 speech database \cite{Schroder} is analyzed. This database contains three voice qualities (modal, soft and loud) uttered by a German female speaker, with about 50 minutes of speech available for each voice quality.

% Note that the conclusions of these latter three sections will be turned to account in Section \ref{sec:VQmodif} for modifying the expressivity in HMM-based speech synthesis.

% In Section \ref{ssec:glottal}, the glottal flow estimated by the complex cepstrum-based decomposition \cite{CC} is extracted and differences in its parametrization are underlined.

%Section \ref{ssec:Eigen} shows some differences between the so-called \emph{eigenresiduals} \cite{DSM} of the different voice qualities.

\subsection{Glottal source}\label{ssec:glottal}

We recently showed that complex cepstrum can be efficiently used for glottal flow estimation \cite{CC}. This method aims at separating the minimum and maximum-phase components of the speech signal. Indeed it has been shown previously \cite{Mixed-phase} that speech is a mixed-phase signal where the maximum-phase (i.e anti-causal) contribution corresponds to the glottal open phase, while the minimum-phase component is related to the vocal tract transmittance (assuming an abrupt glottal return phase). Isolating the maximum-phase component of speech then provides a reliable estimation of the glottal source, which can be achieved by the complex cepstrum. The glottal flow open phase is then parametrized by three features: the glottal formant frequency ($F_g$), the Normalized Amplitude Quotient ($NAQ$) and the Quasi-Open Quotient ($QOQ$). 

\begin{figure*}[!ht]
  \centering
  \includegraphics[width=0.95\textwidth]{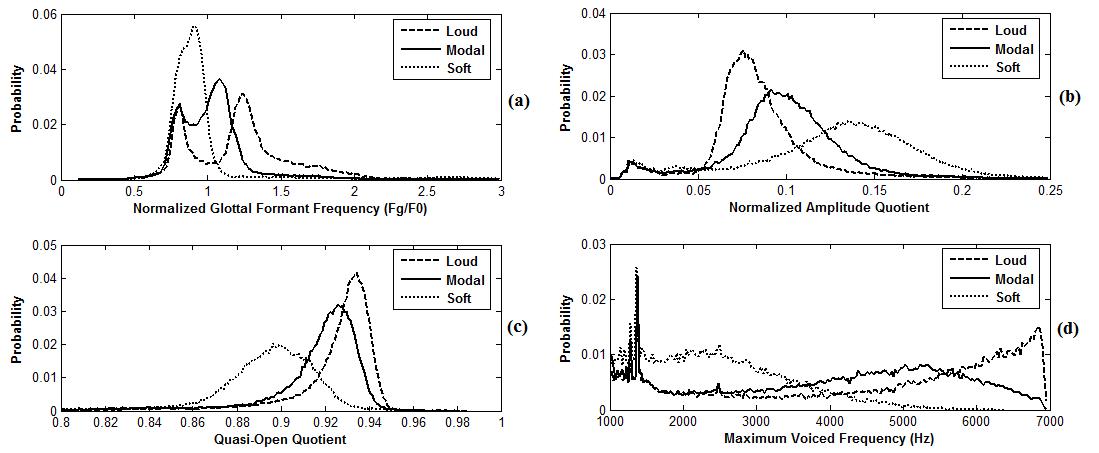}
  \caption{Histograms, for the three voice qualities, of \emph{(a):} the normalized glottal formant frequency $F_g/F_0$, \emph{(b):} the Normalized Amplitude Quotient $NAQ$, \emph{(c):} the Quasi-Open Quotient $QOQ$, and \emph{(d):} the maximum voiced frequency $F_m$.}
  \label{fig:TotalHisto}  
\end{figure*}

The glottal formant was tracked using the method described in \cite{Bozkurt-Fg}. Figure \ref{fig:TotalHisto}(a) shows the histograms of $Fg/F_0$ for the three voice qualities. Significant differences between the distributions are observed. Among others it turns out that a louder (softer) voice results in the production of a higher (lower) glottal formant frequency. Another observation that can be drawn from this figure is the presence of two modes for the modal and loud voices. This may be explained by the fact that the estimated glottal source sometimes comprises a ripple both in the time and frequency domains, which in turn may have two possible causes: an incomplete separation between $Fg$ and the first formant $F_1$ \cite{Bozkurt-Fg}, and/or a non-linear interaction between the vocal tract and the glottis \cite{Plumpe}. This ripple may therefore affect the detection of the glottal formant frequency and in this way explain the parasitical peak in the $Fg/F_0$ histogram for the modal and loud voices.

%\begin{figure}[!ht]
%  % Requires \usepackage{graphicx}
%  \centering
%  \includegraphics[width=0.45\textwidth]{HistoFg.jpg}
%  \caption{Histograms of the glottal formant frequency $F_g$ for the three voice qualities.}
%  \label{fig:HistoFg}  
%\end{figure}

In previous works \cite{Alku-NAQ}, \cite{Alku-QOQ}, Alku et al. proposed the Normalized Amplitude Quotient and the Quasi-Open Quotient as two efficient time-domain parameters characterizing respectively the closing and open phase of the glottal flow. These parameters are extracted using the Aparat toolkit \cite{Aparat} from the glottal source estimated here by the complex cepstrum . Figures \ref{fig:TotalHisto}(b) and \ref{fig:TotalHisto}(c) display the histograms of these two features for the three voice qualities. Notable differences between histograms may be observed.

%\begin{figure}[!ht]
%  % Requires \usepackage{graphicx}
%  \centering
%  \includegraphics[width=0.45\textwidth]{HistoNAQ.jpg}
%  \caption{Histograms of the Normalized Amplitude Quotient $NAQ$ for the three voice qualities.}
%  \label{fig:HistoNAQ}  
%\end{figure}

%\begin{figure}[!ht]
%  % Requires \usepackage{graphicx}
%  \centering
%  \includegraphics[width=0.45\textwidth]{HistoQOQ.jpg}
%  \caption{Histograms of the Quasi-Open Quotient $QOQ$ for the three voice qualities.}
%  \label{fig:HistoQOQ}  
%\end{figure}

\subsection{Maximum Voiced Frequency}\label{ssec:Fm}

Some approaches, such as the Harmonic plus Noise Model (HNM ,\cite{HNM}), consider that the speech signal can be modeled by a non-periodic component beyond a given frequency. In the case of HNM, this maximum voiced frequency ($F_m$) demarcates the boundary between two distinct spectral bands, where respectively an harmonic and a stochastic modeling are supposed to hold. The higher the $F_m$, the stronger the harmonicity, and consequently the weaker the presence of noise in speech. In this paper, $F_m$ was estimated using the algorithm described in \cite{HNM}. Figure \ref{fig:TotalHisto}(d) displays the histograms of $F_m$ for the three voice qualities. It can be observed that, in general, the soft voice has a low $F_m$ (as a result of its breathy nature) and that the stronger the vocal effort, the more harmonic the speech signal, and consequently the higher $F_m$.

%\begin{figure}[!ht]
%  % Requires \usepackage{graphicx}
%  \centering
%  \includegraphics[width=0.45\textwidth]{HistoFm.jpg}
%  \caption{Histograms of the maximum voiced frequency $F_m$ for the three voice qualities.}
%  \label{fig:HistoFm}  
%\end{figure}

\subsection{Spectral Tilt}\label{ssec:SpectralTilt}

\begin{figure}[!ht]
  \centering
  \includegraphics[width=0.45\textwidth]{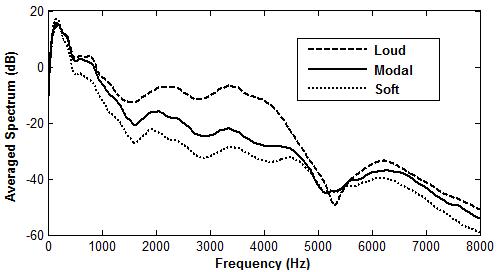}
  \caption{Averaged spectrum for the three voice qualities.}
  \label{fig:AvSpec}
  \vspace{-12pt}    
\end{figure}

Spectral tilt of speech is known to play an important role in the perception of a voice quality \cite{Turk}. To capture this crucial feature, an averaged spectrum is obtained on the whole corpus by a process independent of the prosody and the vocal tract variations. For this, voiced speech frames are extracted by applying a two pitch period-long Hanning windowing centered on the current Glottal Closure Instant (GCI). GCI locations are determined using the technique described in \cite{Drugman-GCI}. These frames are then resampled on a fixed number of points (corresponding to two mean pitch periods) and normalized in energy. The averaged spectrum is finally achieved by averaging the spectra of the normalized speech frames. The averaged amplitude spectrum then contains a mix of the average glottal and vocal tract contributions. The averaged spectrum for the three voice qualities is exhibited in Figure \ref{fig:AvSpec}. Since these spectra were computed for the same speaker, it is reasonable to think that the main difference between them is due to the spectral tilt regarding the produced voice quality. Among others it can be noticed from Figure \ref{fig:AvSpec} that the stronger the vocal effort, the richer the spectral content in the [1kHz-5kHz] band.

% This last step ensures the process to be prosody-independent

% Since the corpus is phonetically balanced, it can be assumed that the vocal tract variations are removed as the formants statistically cancel each other.

\subsection{Eigenresiduals}\label{ssec:Eigen}

\begin{figure}[!ht]
  \centering
  \includegraphics[width=0.45\textwidth]{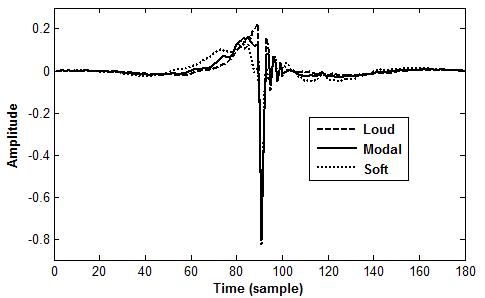}
  \caption{First eigenresidual for the three voice qualities.}
  \label{fig:Eigenresiduals}  
  \vspace{-12pt}
\end{figure}

In \cite{DSM} we proposed to model the residual signal by decomposing speaker-dependent pitch-synchronous residual frames on an orthonormal basis. It was also shown that the first so-obtained eigenvector (called \emph{eigenresidual}) can be efficiently used in parametric speech synthesis. As eigenresiduals are employed in our voice quality modification application in Section \ref{sec:VQmodif}, Figure \ref{fig:Eigenresiduals} displays the differences present in this signal depending on the produced voice quality. It can be noticed that the conclusions we drew in Section \ref{ssec:glottal} about the glottal open phase are corroborated. It is indeed observed that the stronger the vocal effort, the faster the response of the eigenresidual open phase.

% The residual signal obtained by inverse filtering is known to be informative about the speaker identity \cite{Yegna}.
% In \cite{DSM} we modeled this signal by decomposing speaker-dependent pitch-synchronous residual frames on an orthonormal basis.

\subsection{Separability between Distributions}\label{ssec:Separability}

Important differences in the distributions of the features have been presented in the previous subsections (which are in line with the conclusions presented in various studies \cite{Klatt}, \cite{Alku-NAQ}, \cite{Turk}). In this section, we quantify how these differences between voice qualities are significative. For this, the Kullback-Leibler (KL) divergence \cite{KL} is known to measure the separability between two discrete density functions $A$ and $B$. But since this measure is non-symmetric (and consequently is not a true distance), its symmetrised version, called Jensen-Shannon divergence \cite{KL}, is often prefered. It consists of a sum of two KL measures:

\vspace{-8pt}
\begin{equation*}\label{eq:DJS}
D_{JS}(A,B)=\frac{1}{2}(\sum_i{A(i)\log_2 \frac{A(i)}{M(i)}}+\sum_i{B(i)\log_2 \frac{B(i)}{M(i)}})
\end{equation*}

%\begin{equation}\label{eq:DKL}
%D_{KL}(A,B)=\sum_i{A(i)\log_2 \frac{A(i)}{B(i)}}
%\end{equation}

%\begin{equation}\label{eq:DJS}
%D_{JS}(A,B)=\frac{1}{2}D_{KL}(A,M)+\frac{1}{2}D_{KL}(B,M)
%\end{equation}

where $M$ is the average of the two distributions ($M=0.5*(A+B)$). Table \ref{tab:TabDJS} shows the results for the four features we previously presented. Among others it can be noticed that the loud and soft voices are highly separable, while the loud type is closer to the modal voice than the soft one. It is also seen that $F_g$ and $NAQ$ are highly informative for voice quality labeling.

\begin{table}[!ht]
\centering
\begin{tabular}{l | c | c | c | c|}
  & $F_g$ & $NAQ$ & $QOQ$ & $F_m$ \\
\hline
$D_{JS}(Loud,Modal)$ & 0.196 & 0.118 & 0.035 & 0.076\\
\hline
$D_{JS}(Loud,Soft)$ & 0.353 & 0.371 & 0.279 & 0.297\\
\hline
$D_{JS}(Modal,Soft)$ & 0.175 & 0.194 & 0.175 & 0.215\\
\hline
\end{tabular}
\caption{Jensen-Shannon divergence between the three voice qualities for the four extracted features.}
\label{tab:TabDJS}
\end{table}

%%%%%%%%%%
%%%%%%%%%%
\section{Voice quality modification}\label{sec:VQmodif}

In a previous work \cite{DSM}, we proposed a Deterministic plus Stochastic Model (DSM) of the residual signal. In this approach, the excitation is divided into two distinct spectral bands delimited by the maximum voiced frequency $F_m$. The deterministic part concerns the low-frequency contents and is modeled by the first eigenresidual as explained in Section \ref{ssec:Eigen}. As for the stochastic component, it is a high-pass filtered noise similarly to what is used in the HNM \cite{HNM}. The residual signal is then passed through a LPC-like filter to obtain the synthetic speech.
This section aims at applying voice quality modification as a post-process to HMM-based speech synthesis \cite{HTS} using the DSM of the residual signal. More precisely, a HMM-based synthesizer is trained on a corpus of modal voice for a given speaker. The goal is then to transform the synthetic voice so that it is perceived as soft or loud, while avoiding a degradation of quality in the produced speech.

Since no dataset of expressive voice is available for the considered speaker, modifications are extrapolated from the prototypes described for speaker De7 in Section \ref{sec:VQanalysis}, assuming that other speakers modify their voice quality in the same way. Three main transformations are here considered:

\begin{itemize}
\item The eigenresiduals presented in Section \ref{ssec:Eigen} are used for the deterministic part of the DSM. These waveforms implicitly convey the modifications of glottal open phase that were underlined in Section \ref{ssec:glottal}.

\item The maximum voiced frequency $F_m$ is fixed for a given voice quality according to Section \ref{ssec:Fm} by taking its mean value: 4600 Hz for the loud, 3990 Hz for the modal (confirming the 4 kHz we used in \cite{DSM}), and 2460 Hz for the soft voice.

\item The spectral tilt is modified using the inverse of the process described in Section \ref{ssec:SpectralTilt}. For this, the averaged spectrum of the voiced segments is transformed, in the pitch-normalized domain, by a filter expressed as a ratio between auto-regressive modelings of the source and target voice qualities (cf Fig.\ref{fig:AvSpec}). Residual frames are then resampled to the target pitch at synthesis time. This latter transformation is then pitch-dependent.
\end{itemize}

To evaluate the technique, a subjective test was submitted to 10 people. The test consisted of 27 sentences generated by our system for three speakers (two males and one female). One third of these sentences were converted to a softer voice, and one third to a louder one. For each sentence, participants were asked to assess the vocal effort they perceive (0 = very soft, 100 = very tensed), and to give a MOS score. Results are displayed in Table \ref{tab:ResultTest} with their 95\% confidence intervals. Interestingly it can be noticed that voice quality modifications are perceived as expected while the overall quality is not significantly altered (although listeners have a natural tendency to prefer softer voices).

\begin{table}[!ht]
\centering
\begin{tabular}{c | c | c |}
  & Effort ratings & MOS scores \\
\hline
Modal to Soft & 36.11 $\pm$ 2.60  & 3.189 $\pm$ 0.145  \\
\hline
Modal & 52.89 $\pm$ 2.82 & 3.017 $\pm$ 0.147  \\
\hline
Modal to Loud & 72.11 $\pm$ 2.60  & 2.606 $\pm$ 0.146  \\
\hline
\end{tabular}
\caption{Perceived vocal effort ratings (0 = very soft voice, 100 = very tensed voice) and MOS scores for the three versions together with their 95\% confidence intervals.}
\label{tab:ResultTest}
\vspace{-12pt}
\end{table}

%\begin{figure*}[!ht]
%  % Requires \usepackage{graphicx}
%  \centering
%  \includegraphics[width=0.95\textwidth]{ResultModif.jpg}
%  \caption{\emph{Left: }MOS scores for the three versions, \emph{Right: }Perceived vocal effort for the three versions (0 = very soft voice, 100 = %very tensed voice).}
%  \label{fig:ResultModif}  
%\end{figure*}

%\begin{figure}[!ht]
%  % Requires \usepackage{graphicx}
%  \centering
%  \includegraphics[width=0.45\textwidth]{VocalEffort.jpg}
%  \caption{Perceived vocal effort for the three versions (0 = very soft voice, 100 = very tensed voice).}
%  \label{fig:VocalEffort}  
%\end{figure}

%\begin{figure}[!ht]
%  % Requires \usepackage{graphicx}
%  \centering
%  \includegraphics[width=0.45\textwidth]{MOS.jpg}
%  \caption{MOS scores for the three versions.}
%  \label{fig:MOS}  
%\end{figure}

%%%%%%%%%%
%%%%%%%%%%
\section{Conclusion}\label{sec:conclu}

In this study we show that a glottal flow estimation algorithm \cite{CC} can be effectively used for voice quality analysis on large speech corpora where most of glottal flow estimation literature are based on tests with sustained vowels. We study the variations in parameters for different voice qualities and conclude that the two glottal flow parameters $F_g$ and $NAQ$ are highly informative for voice quality labeling. We further show that the information extracted from one speech database can be applied to other speech databases for voice quality modification and the quality achieved in a speech synthesis application is fairly high.

%%%%%%%%%%
%%%%%%%%%%
\section{Acknowledgments}\label{sec:Acknowledgments}

Thomas Drugman is supported by the ``Fonds National de la Recherche
Scientifique'' (FNRS). The authors would like to thank M. Schroeder for the De7 database, as well as Y. Stylianou for providing the algorithm extracting $F_m$.

\bibliographystyle{IEEEbib}
\bibliography{strings,refs}

\end{document}